\documentclass[onecolumn,sort&compress,numbers]{els-mrw} 

\usepackage{amsmath,amssymb,amsfonts,amsthm,makeidx,graphicx}
\usepackage{txfonts}
\usepackage{helvet}
\usepackage{siunitx}
\usepackage{subfigure}

\begin{document}


\chapter{Transforming Antarctic Ice into a Cherenkov Neutrino Detector}\label{chap1}

\author[1]{Francis Halzen}%
\author[1]{John Kelley}%

\address[1]{\orgname{University of Wisconsin--Madison}, \orgdiv{Wisconsin IceCube Particle Astrophysics Center}, \orgaddress{222 W. Washington Ave., Suite 500, Madison WI 53703}}

\articletag{Chapter Article tagline: update of previous edition, reprint.}

\maketitle

\begin{abstract}[Abstract]
In this chapter, we describe how the IceCube Neutrino Observatory transformed a cubic kilometer of natural ice at the geographic South Pole into a neutrino telescope. The concept of using the neutrino as an astronomical messenger is as old as the neutrino itself, and the challenge to open this new window on the high-energy universe was technological in nature. We discuss how IceCube was constructed and how the detector operates, including some detail on the 5,484 optical sensors that comprise the array. We highlight some of the primary results of the experiment, including the discovery of a diffuse flux of high-energy neutrinos reaching us from the cosmos, the observation of the first high-energy neutrino sources in the sky, and the observation of our Galaxy in neutrinos.
\end{abstract}

\begin{keywords}
 	neutrino detector \sep particle astrophysics \sep Cherenkov \sep Antarctica \sep ice
\end{keywords}
\begin{figure}[ht]
	\centering
	\includegraphics[height=12cm]{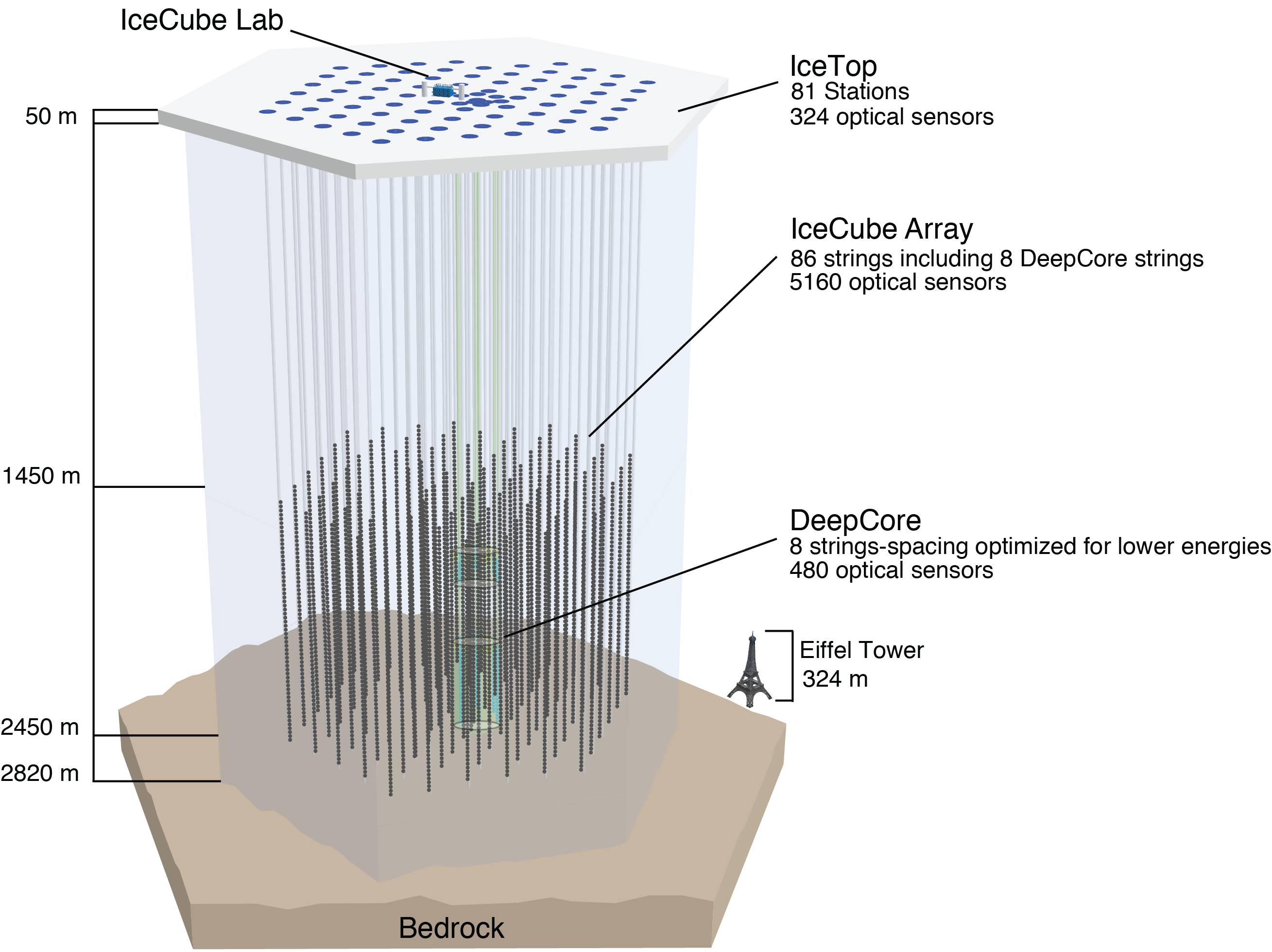}
	\caption{The IceCube Neutrino Observatory transforms a cubic kilometer of ice at the South Pole into a neutrino detector.}
	\label{fig:titlepage}
\end{figure}


\section*{Objectives}
\begin{itemize}
	\item Describe the particle physics and astrophysics motivating 
      neutrino astronomy and the instrumentation required; 
	\item explain how the IceCube neutrino detector works; and
	\item highlight the primary physics results of IceCube and the next steps in neutrino astronomy.
\end{itemize}





\section{An Introduction to Neutrino Astronomy: Instrumentation and Methods}\label{generic}

The IceCube Neutrino Observatory is the first astronomical telescope of its kind, designed to observe the cosmos with neutrinos rather than light. 
This “telescope” is a Cherenkov detector built deep inside the Antarctic ice sheet that covers the geographic South Pole. Encompassing a cubic kilometer of ice located 1.5 kilometers below the surface (see Fig.~\ref{fig:titlepage}), IceCube searches for subatomic particles called neutrinos reaching us from the cosmos. These astronomical messengers provide the information to unlock the secrets of high-energy cosmic particle accelerators powered by supermassive black holes and create the opportunity to study neutrinos at a million times the energy of those produced at laboratory particle accelerators such as Fermilab.

Neutrinos are often referred to as ghost particles because they cannot be observed directly. Rarely, about one in a million for the neutrinos passing through IceCube, a neutrino will crash into a molecule of ice and trigger a small nuclear explosion that produces a flash of blue light. The light pattern is viewed by IceCube’s more than 5,000 sensors embedded in the ice, which is both the Cherenkov medium and the support structure for the experiment. It reveals the energy of the neutrino and its direction, which pinpoints where it originated in the universe. Hence, this particle detector is an astronomical telescope.

\subsection{Motivation}

The shortest wavelength radiation reaching us from the universe is not radiation at all; it consists of cosmic rays --- high-energy nuclei, mostly protons. Some reach us with extreme energies exceeding $10^8$~TeV from a universe beyond our Galaxy that is obscured to gamma rays and from which only neutrinos reach us as astronomical messengers~\cite{Aartsen:2013jdh} that point back to the sources where they originate. The recent identification of neutrinos produced in the dense core surrounding the central supermassive black hole of active galaxies~\cite{IceCube:2018dnn,IceCube:2018cha,IceCube:2022der} represents a breakthrough towards resolving the century-old mystery of the origin of cosmic rays. Where cosmic rays originate in our own Galaxy remains a puzzle, although the discovery that our Galaxy emits neutrinos may represent a promising path for progress~\cite{IceCube:2023ame}.

\begin{figure}[ht]
\centering
\includegraphics[width=0.4\linewidth]{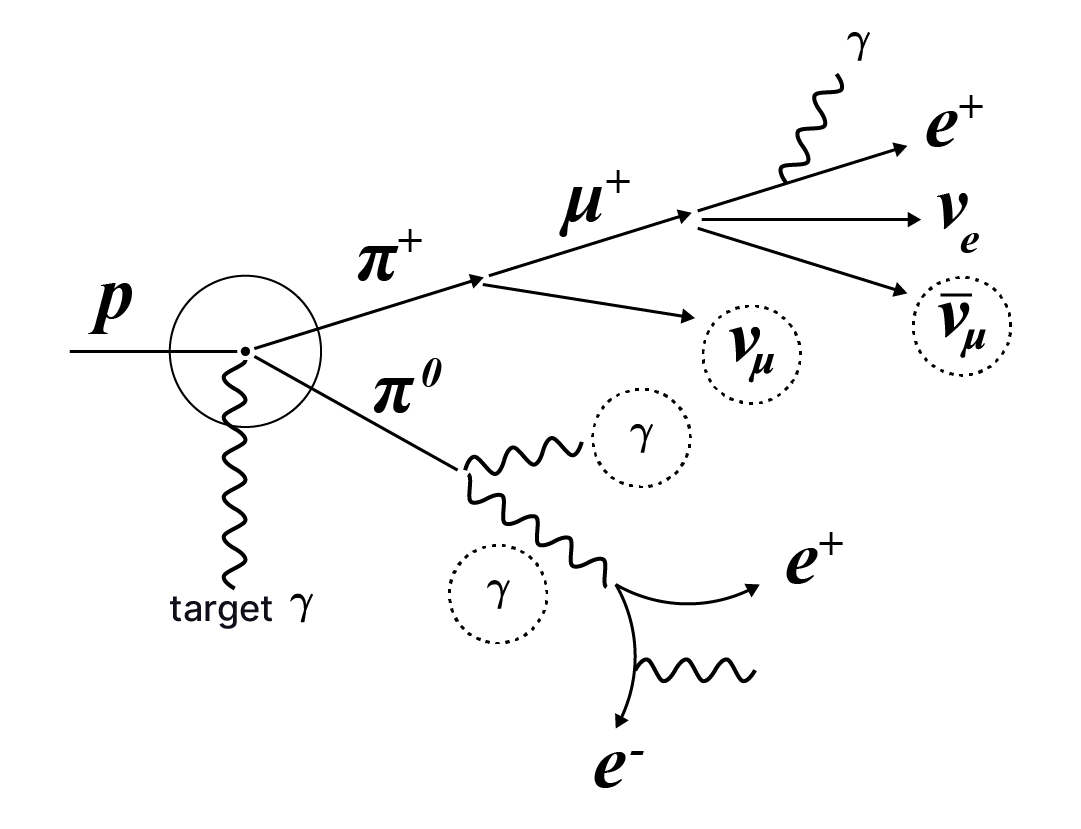}
\caption{Flow diagram showing the production of charged and neutral pions in $p\gamma$ interactions. The circles indicate equal energy going into gamma rays and into pairs of muon neutrinos and antineutrinos, which IceCube cannot distinguish between. Because the charged pion energy is shared roughly equally among four particles and the neutral pion energy among two photons, the photons have twice the energy of the neutrinos.}
\label{fig:flow}
\end{figure}

It is generally agreed that the enormous energy and luminosity of cosmic-ray accelerators can only be powered by the gravitational energy of neutron stars and black holes. The rationale for searching for cosmic-ray sources by observing neutrinos is straightforward: in relativistic particle flows onto neutron stars or black holes, some of the gravitational energy released in the accretion of matter is transformed into the acceleration of protons or heavier nuclei, which subsequently interact with radiation and/or ambient matter to produce pions and other secondary particles that decay into neutrinos. For instance, when accelerated protons interact with the intense radiation fields surrounding the black hole via the photoproduction processes
\begin{equation}
p + \gamma \rightarrow \pi^0 + p
\mbox{ \ and \ }
p + \gamma \rightarrow \pi^+ + n\,,
\label{eq:delta}
\end{equation}
both neutrinos and gamma rays are produced with roughly equal rates; while neutral pions decay into two gamma rays, $\pi^0\to\gamma+\gamma$, the charged pions decay into three high-energy neutrinos ($\nu$) and antineutrinos ($\bar\nu$) via the decay chain $\pi^+\to\mu^++\nu_\mu$ followed by $\mu^+\to e^++\bar\nu_\mu +\nu_e$; see Fig.~\ref{fig:flow}. Based on this simplified flow diagram, we expect equal fluxes of gamma rays and pairs of $\nu_\mu+\bar\nu_\mu$ neutrinos, which a neutrino telescope cannot distinguish between. Also, from the fact that in the photoproduction process 20\% of the initial proton energy is transferred to the pion, we anticipate that the gamma ray carries one tenth of the proton energy and the neutrino approximately half of that. Because cosmic neutrinos are inevitably accompanied by high-energy photons, neutrino astronomy is a multimessenger astronomy. Importantly, neutrinos trace sources where protons are accelerated to produce pions, i.e., cosmic-ray sources.

The actual relationship between photon and neutrino emission in multimessenger sources is unlikely to be that simple. First, one must separate photons associated with neutrinos, which we will refer to as {\em pionic} photons, from photons radiated by electrons that are accelerated along with cosmic rays. Second, the pionic photons do not reach our telescopes with their initial energy. They interact in the extragalactic background light (EBL) at the highest energies, predominantly in interactions with microwave photons, $411\ \rm cm^{-3}$, via the process $\gamma + \gamma_{\it cmb} \rightarrow e^+ + e^-$, initiating an electromagnetic cascade that distributes their initial energy over multiple gamma rays that arrive at Earth with energies within the sensitivity of the NASA Fermi Gamma-ray Space Telescope (Fermi). Additionally, powerful neutrino sources within reach of IceCube's sensitivity are likely to require an efficient --- i.e., dense --- target for producing neutrinos that is likely to be opaque to the accompanying pionic gamma rays. Because of the additional energy loss in the source, they may appear below the threshold of Fermi and ground-based atmospheric gamma-ray telescopes, with the energy in pionic photons emerging at MeV energies and below; we will refer to them as {\it gamma-obscured} sources. IceCube's observation of gamma-obscured sources has confirmed this expectation.

\subsection{Detecting neutrinos}

The idea that the neutrino, discovered in 1956, represents an ideal astronomical messenger is as old as the neutrino itself. As 
early as the late 1970s, estimates suggested that a kilometer-scale detector may be required to open a neutrino window on the 
universe. The method for building such an instrument had already been presented by Moshe Markov at the first Rochester 
conference in 1960. After that, it became a technological challenge.

\begin{figure}
\centering
\begin{subfigure}
\centering
\includegraphics[width=.8\textwidth]{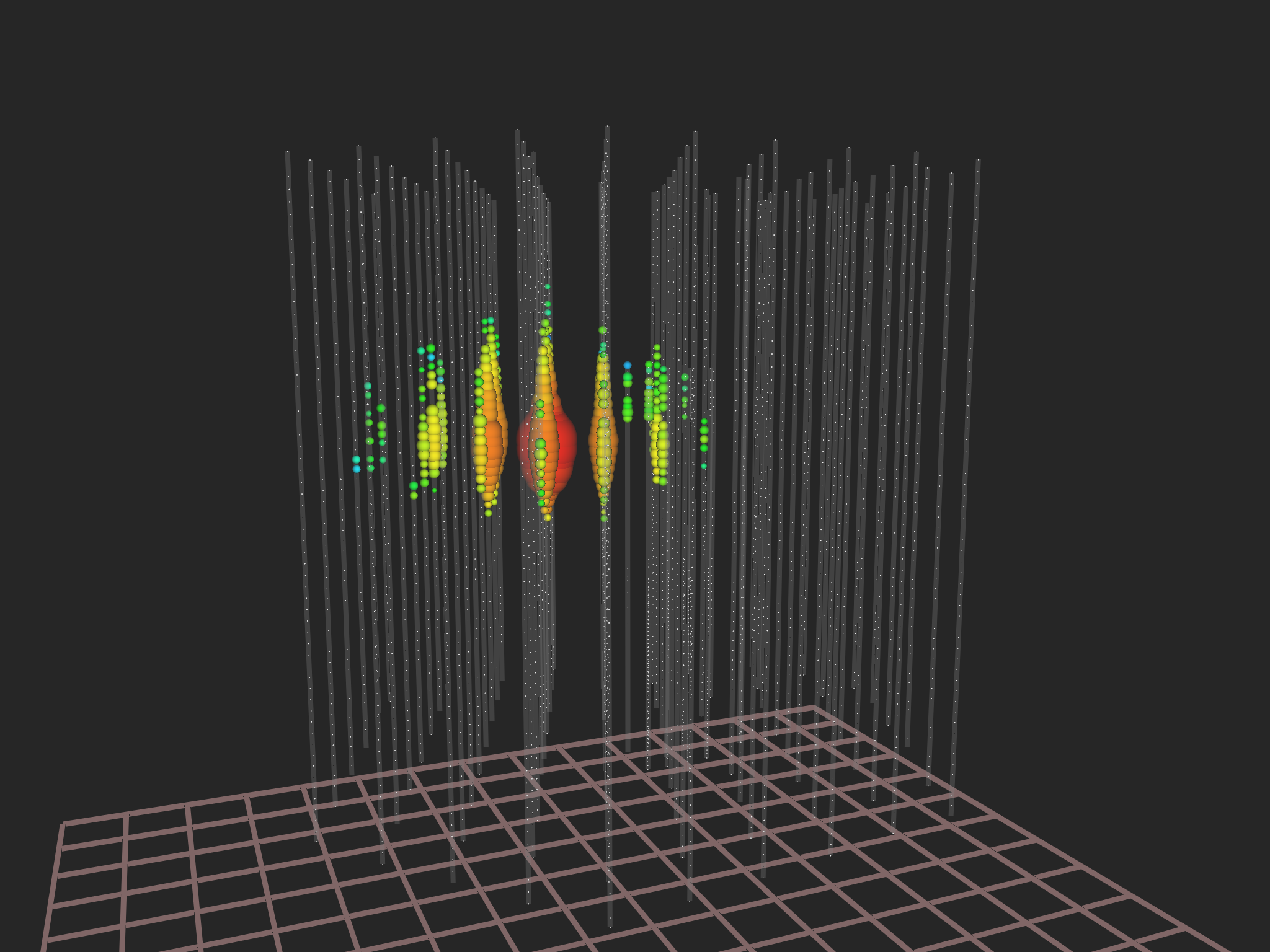}
\end{subfigure}
\begin{subfigure}
\centering
\includegraphics[width=.8\textwidth]{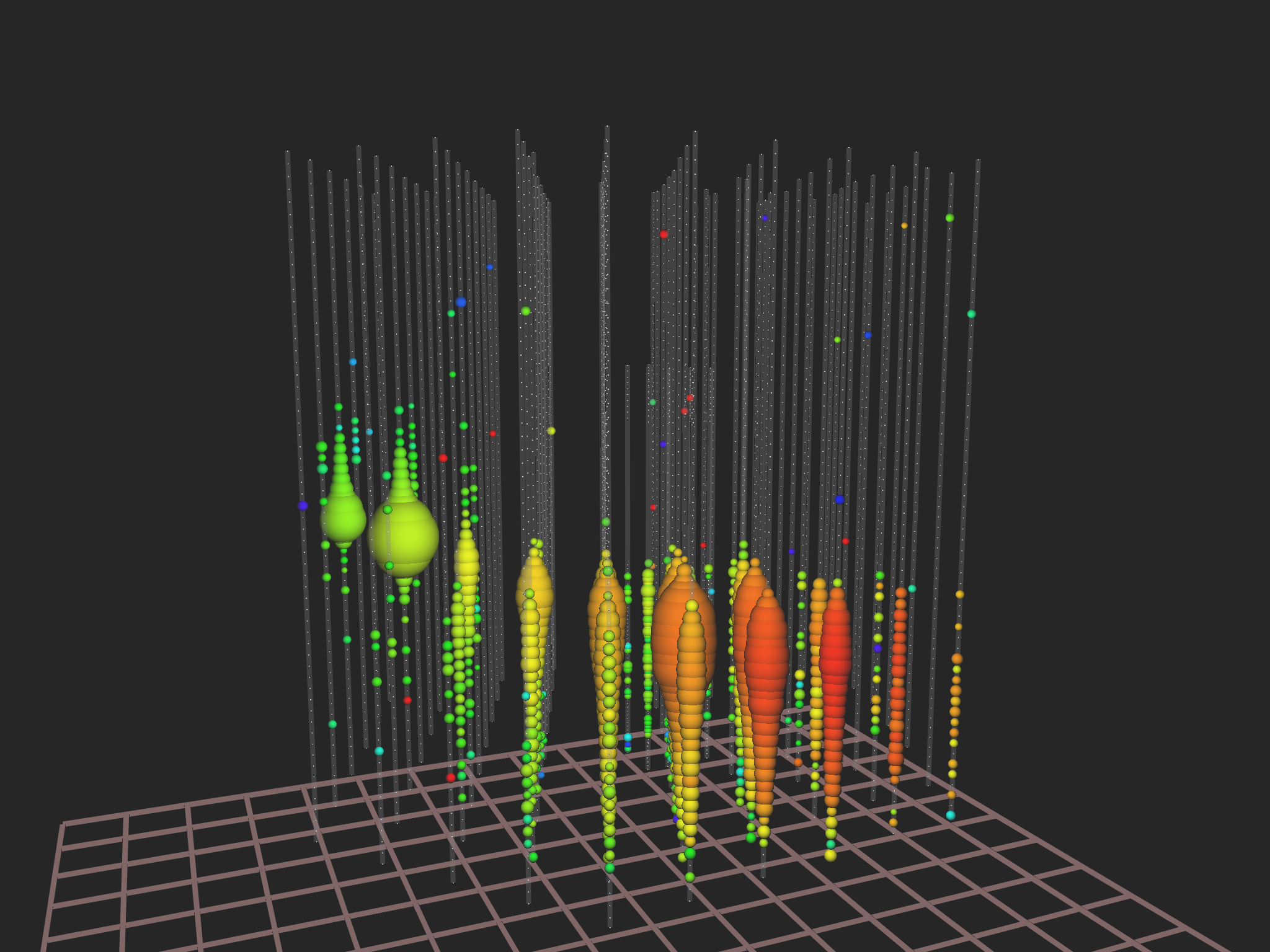}
\end{subfigure}
\caption{{\bf Top Panel:}  Light pattern produced in IceCube by a shower initiated by an electron or tau neutrino of $1.14$ PeV. Tiny white dots along the vertical strings represent sensors with no signal. The color of the dots indicates the arrival time of the Cherenkov photons, from red (early) to purple (late), following the rainbow, and their size reflects the number of photons detected. {\bf Bottom Panel:}  A muon track coming up through the Earth traverses the detector at an angle of $11^\circ$ below the horizon. The deposited energy, i.e., the energy equivalent of the total Cherenkov light of all charged secondary particles inside the detector, is 2.6\,PeV .}.
\label{fig:erniekloppo}
\end{figure}

Located near the National Science Foundation's research station at the geographical South Pole, the IceCube project~\cite{Aartsen:2016nxy} first met the challenge by transforming one cubic kilometer of natural Antarctic ice into a Cherenkov detector. Below a depth of 1,450\,meters, the glacial ice has been instrumented with 86 cables called ``strings,'' each of which is equipped with 60 optical sensors. Each sensor, or digital optical module (DOM), consists of a glass sphere containing a 10-inch photomultiplier tube (PMT) and the electronics board that captures and digitizes the signals locally using an onboard computer. The digitized signals are given a time stamp with residuals accurate to 2\,nanoseconds and are subsequently transmitted to the surface. Processors at the surface continuously collect the time-stamped signals from the optical modules, each of which functions independently. The digital messages are sent to a string processor and a global event trigger. The events contain the Cherenkov radiation patterns resulting from either the catastrophic energy losses of a secondary muon track produced by a muon neutrino interacting in the ice, or by secondary particle showers for the case of electron and tau neutrinos --- ``tracks" or ``cascades"; see Fig.~\ref{fig:erniekloppo}. These reveal the flavor, energy, and direction of the incident neutrino. 

Neutrinos interacting inside or in the vicinity of the detector produce leptons of the corresponding neutrino flavor carrying 80\% of their initial energy at the highest energies. The properties of the neutrino are inferred from the Cherenkov radiation pattern produced by the secondary charged particles and recorded by the optical modules. The arrival direction of a secondary muon track or of an electromagnetic shower initiated by an electron or tau neutrino is determined by the arrival times of the Cherenkov photons at the optical sensors, while the number of Cherenkov photons stands as a proxy for the energy deposited by secondary particles in the detector. Although the detector only records the energy of the secondary muon inside the detector, from Standard Model physics we can infer the likely energy of the parent neutrino; for a more detailed discussion, see Ref.~\cite{Halzen:2006mq}.

After a decade of refining the geometry and calibration of the instrument and optimizing the analysis techniques using improved modeling of the optics of the ice, muon tracks resulting from muon neutrino interactions can be pointed back to their sources with a $\le 0.3^\circ$ angular resolution for high-energy events. In contrast, the reconstruction of cascade directions --- in principle, possible to within a few degrees --- is still in the development stage in IceCube, achieving $5^\circ$ resolution~\cite{Tyuan2017}. However, determining their energy from the observed light deposition is straightforward, and a resolution of better than 15\% can be achieved. For illustration, we contrast in Fig.~\ref{fig:erniekloppo} the Cherenkov patterns initiated by an electron (or tau) neutrino of 1\,PeV energy (top) and a neutrino-induced muon track depositing 2.6\,PeV energy in the detector (bottom).

\subsection{Construction}

Bringing AMANDA, the R\&D project for IceCube, and IceCube itself to fruition presented three main challenges: developing the drilling system to position instrumentation at depth, determining the optical properties of natural Antarctic ice, and rejecting large backgrounds of muons and neutrinos produced in the Earth's atmosphere in order to reach sensitivity to the low fluxes of cosmic neutrinos anticipated but certainly not guaranteed. Already by the 1970s, it had been understood~\cite{Roberts:1992re} that a kilometer-scale detector was needed to observe the so-called GZK neutrinos produced in the interactions of cosmic rays with background microwave photons~\cite{Beresinsky:1969qj, Stecker:1973sy}. This so-called ``guaranteed" flux has actually escaped detection so far because it turned out that the highest energy cosmic rays are heavy nuclei.

The construction season in the Antarctic summer is short, from mid-October through mid-February, and being able to drill holes and deploy strings quickly is critical. At the University of Wisconsin–Madison, we developed a hot water drilling system able to ``drill" holes in less than two days to 2.5 km depth, requiring a power plant of close to 5 megawatts to melt the ice. IceCube DOMs are deployed in water-filled holes, 61 cm in diameter. The water at the edges of these holes begins to refreeze almost immediately; their 61-cm diameter ensures that the holes remain open wide enough to accommodate the cable and DOMs for 30 hours, allowing a full string deployment. 
The ice is melted by a hot water jet under pressure from a nozzle supplying 200 gallons per minute at 6.89 MPa and a temperature of \SI{88}{\celsius}. The water is subsequently pumped out of the hole and returned to a heating system at a rate of 927 L/minute and a temperature of \SI{1}{\celsius}. The water is reheated at the surface and returned to the nozzle at depth. Drilling progresses by circulating this reheated water to ever-increasing depth. The hot water drill reaches 2.5 km depth in less than 30 hours, using less than 30,000 liters of fuel per hole.

\begin{figure}[ht]
\centering
\includegraphics[width=0.5\linewidth]{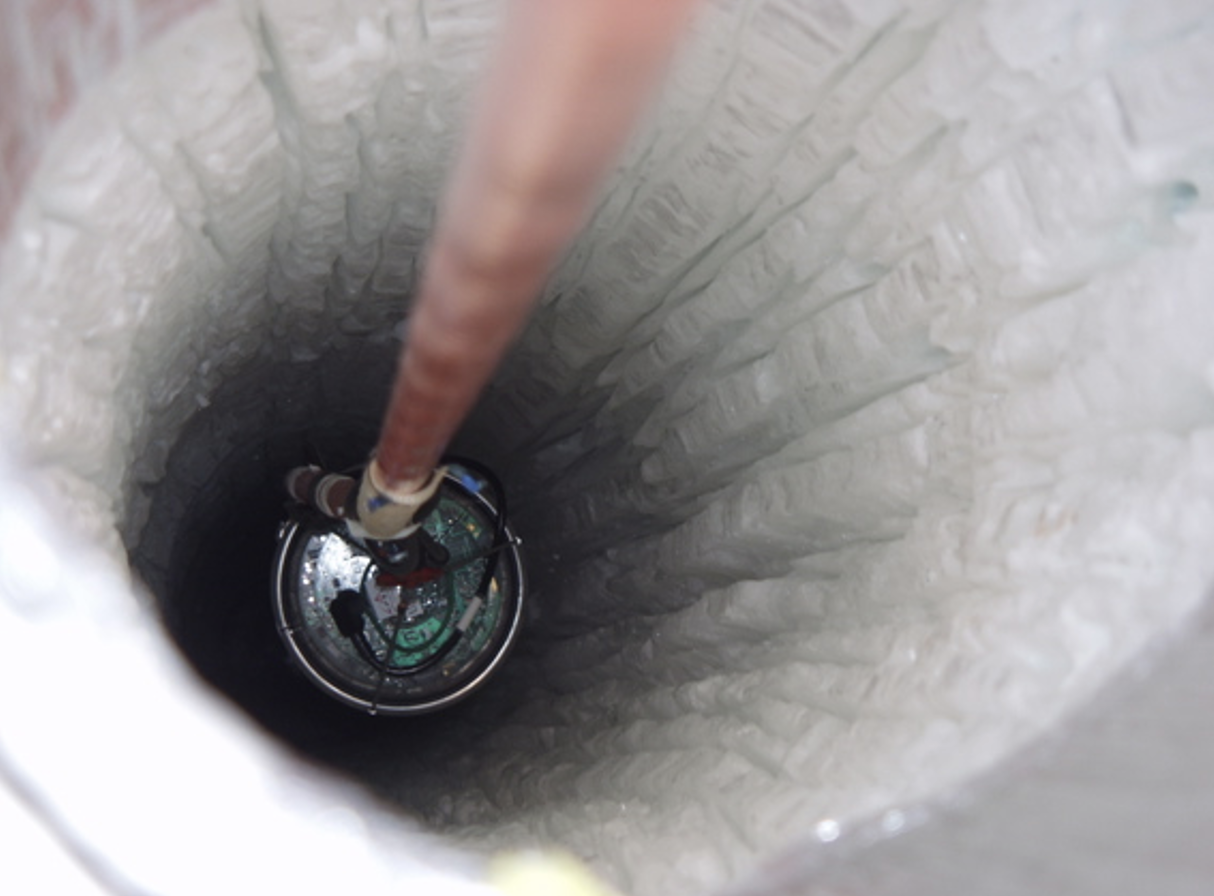}
\caption{An IceCube DOM as it is deployed into a newly drilled hole. \emph{Photo credit: IceCube/NSF}}
\label{fig:dom_in_hole}
\end{figure}

The fuel tanks and all drill components are built on movable sleds to allow for repositioning the drilling infrastructure every new season. The drill speed is computer controlled; the actual width and refreeze time of each hole are accurately computed on the basis of drill data entered into the software. In the 2009--10 season, the system delivered 20 holes with a performance far superior to design. Holes were drilled in as little as 27 hours, with a fuel consumption of just over 15,000 L, greatly improving upon design goals. Before the water in the hole refreezes, a cable is lowered into the hole that will carry the signal to the surface from the 60 DOMs attached every 17 m (see Fig.~\ref{fig:dom_in_hole}). In situ construction of the string and lowering it to depth takes roughly 10 hours. Each string takes data as soon as the hole refreezes.

After establishing that bubbles in the ice completely disappear below \SI{1350}{\meter}, the detector was deployed below \SI{1450}{\meter}. Both the optical absorption and scattering of the photons radiated by charged particles in the ice are important in determining the reconstruction capabilities of IceCube.  The optical transmission depends strongly on impurities in the ice that were introduced when the ice was first laid down as snow at the surface.  This deposition happens in layers; each year the snowfall produces a thin, nearly horizontal layer.  For the ice in IceCube, this happened over the last 100,000 years.  Variations in the long-term dust level in the atmosphere over this period, as well as the occasional volcanic eruption, lead to depth-dependent variations in the absorption and scattering properties that have to be mapped in detail. Much effort has gone into measuring the optical properties of the ice, using artificial light sources and in situ measurements. The studies are made possible by using LEDs and lasers that emit at a variety of wavelengths and are deployed along with the DOMs. By measuring the arrival time distributions of photons at different distances from a light source, it is possible to measure both the attenuation length and scattering length of the light. Higher-resolution depth-dependent measurements of the ice properties are performed by ``dust loggers" that are lowered down the water-filled holes immediately after drilling.  They shine a thin beam of light into the ice and measure the reflected light.  One can thus measure the ice properties with a vertical resolution given by the width of the emitted beam, a few millimeters \cite{icecube2013south}. Other properties of the ice sheet are important and have been measured with increased precision as the data accumulates, such at the deformation of the horizontal layering of the ice, the anisotropy of the scattering relative to the motion of the glacier resulting from the scattering on crystal boundaries and the alignment of the dust particles, and, finally, the alterations of the optics of the refrozen ice after drilling.

 As in conventional astronomy, IceCube looks beyond the atmosphere for cosmic signals. The final challenge has been to establish that cosmic neutrinos can be identified in the large background of cosmic-ray muons and atmospheric neutrinos from the decay of pions and kaons produced in cosmic-ray interactions with nitrogen and oxygen nuclei in the atmosphere. IceCube detects cosmic-ray muons at a rate of $\sim 2,750\ \rm kHz$ and more than 100,000 atmospheric neutrinos per year with energies above a threshold of $100\, \rm GeV$. A subset of these observed with the denser central DeepCore subarray reaches a threshold down to $5 \,\rm GeV$. Atmospheric neutrinos are a background for cosmic neutrinos, at least for energies below 1,000 TeV, but their flux is calculable and can be used to verify the calibration of the detector. At the highest energies, a small charm component is anticipated; its magnitude is uncertain and remains to be measured. The flux of atmospheric neutrinos falls with increasing energy; atmospheric background events above a few 100 TeV are rare, leaving a clear field of view for extraterrestrial sources.

 At lower energies, the downgoing atmospheric neutrino background can be isolated because the neutrinos are accompanied by muons from the shower in which they originated. Traditionally the ratio of signal to background events was further enhanced by quality cuts, for instance on the ratio of the upgoing and downgoing likelihood of the track reconstruction and the number of unscattered Cherenkov photons in the event. After taking a decade of data, improvements to the detector geometry, the calibration of the individual photomultiplier tubes, and the inclusion of the non-Gaussian tails in the point spread function of the muon track reconstruction, a superior agreement between the performance and simulation of the detector has been achieved. This made possible the use of the powerful neural nets that were developed in the meantime, and now neutrino sample selection and their reconstruction are routinely performed using machine learning. For instance, the use of neural nets increased the sample of shower events in the general direction of the Galactic plane by more than an order of magnitude and improved their reconstruction by a factor of two. This made identification of Galactic neutrino emission possible in the presence of the dominant background of neutrinos of extragalactic origin. 


\section{From Data to Physics}\label{data}

\begin{figure}[ht]
\centering
\includegraphics[width=0.7\linewidth]{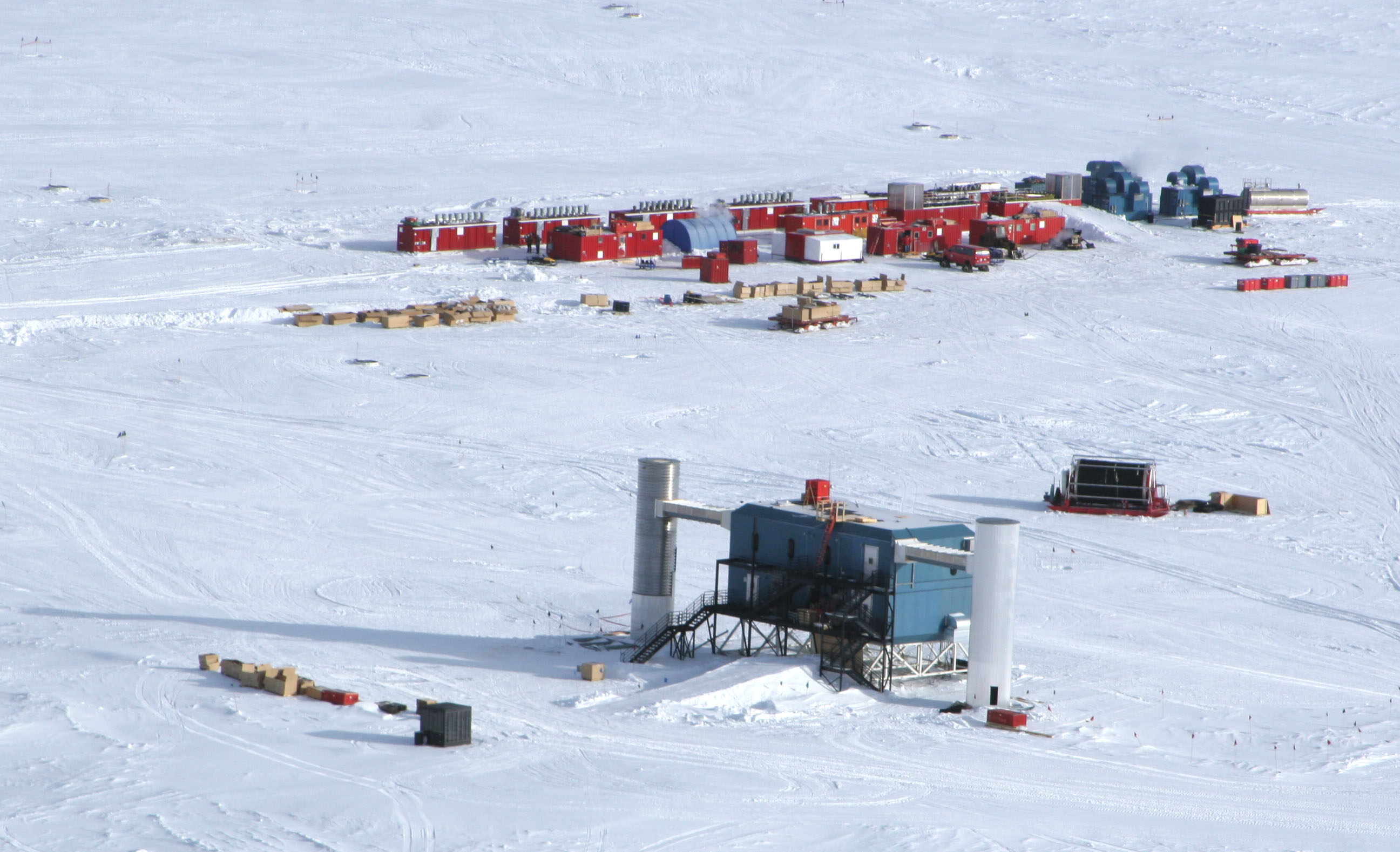}
\caption{The IceCube Laboratory (front, ICL) and drill seasonal equipment site (back, SES) during detector construction. The detector "strings" or cables connect the sensors frozen into the deep glacial ice through the cable towers on each side of the building to readout electronics and computers in the ICL. \emph{Photo credit: E.~Dicks / NSF}}
\label{fig:icl_ses}
\end{figure}

The IceCube detector, shown in Fig.~\ref{fig:titlepage}, consists of 86 in-ice optical detector strings of 60 sensors each, including a dense central subarray ("DeepCore"), and 81 IceTop stations near the surface that each consist of two ice tanks, each of those equipped with two sensors. The cables connect the sensors to the online data acquisition and processing systems in the IceCube Laboratory (ICL, Fig.~\ref{fig:icl_ses}). In the following sections, we describe the detector optical modules and the data flow from the photons in the ice to data processing and analysis.


\subsection{The digital optical module}\label{dom}

\begin{figure}[ht]
\centering
\includegraphics[width=0.4\linewidth]{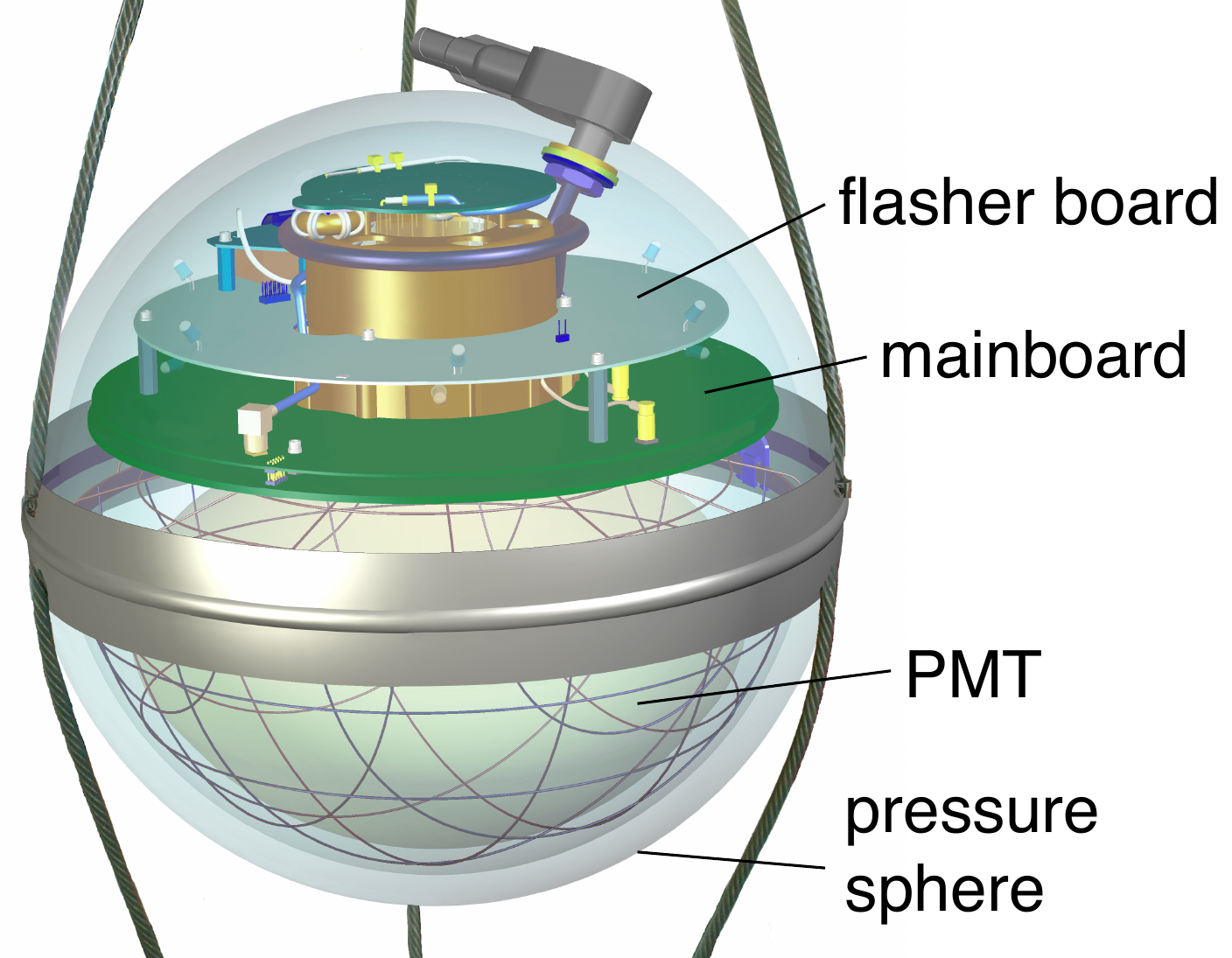}
\caption{Model of the IceCube digital optical module (DOM) with key components.}
\label{fig:dom}
\end{figure}

The primary sensor for IceCube is the digital optical module, or DOM (Fig.~\ref{fig:dom}). The DOMs are responsible for converting the Cherenkov photons emitted by charged particles in the ice into electrical signals and then digitizing these signals for transmission to the IceCube Laboratory at the surface. The timing and amplitude (brightness) information from the DOMs can be used, along with the location of each module in the ice, to reconstruct particle properties such as direction and deposited energy. 

\begin{figure}[ht]
\centering
\includegraphics[width=0.6\linewidth]{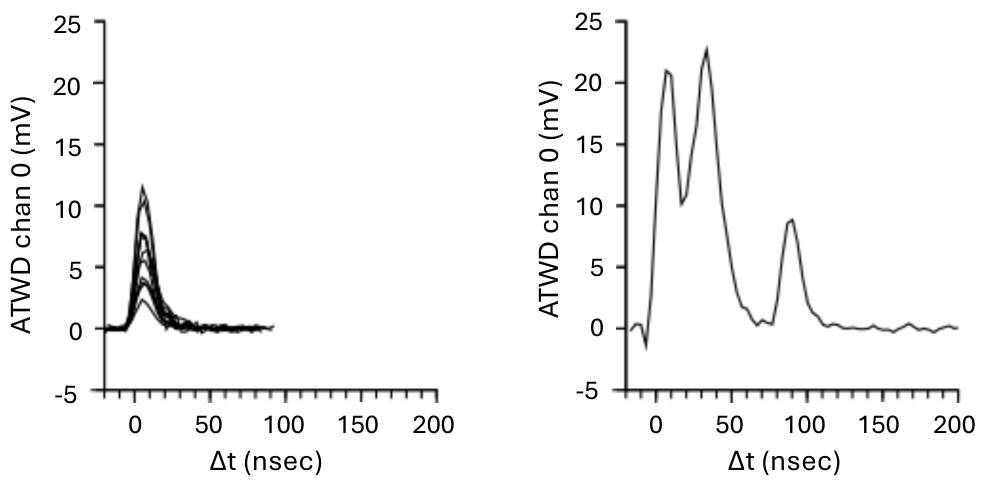}
\caption{\label{fig:pmt_waveforms}Example digitized PMT signals from an IceCube DOM. Left: a collection of SPE waveforms. The spread in amplitudes is due to the stochastic nature of the multiplication process. Right: sample multi-PE waveform.}
\end{figure}

The DOMs detect Cherenkov photons using a 10-inch photomultiplier tube (PMT; see \cite{Abbasi:2010vc}). When a photon hits the coated glass face of the PMT, or photocathode, an electron may be emitted via the photoelectric effect inside the PMT body. A large electric field funnels the electron to a series of plates (dynodes) at intermediate voltages. The initial electron is multiplied into many as it cascades from one dynode to the next; when the ensemble of electrons reach the collecting anode, the charge has been increased by a factor of $10^7$ (the gain of the PMT), which is large enough to record with conventional electronics. The PMT signal from a single photon is known as a single photoelectron, or SPE (Fig.~\ref{fig:pmt_waveforms}). If multiple photons arrive at the PMT, the output signal is the linear combination of SPEs, until the PMT cannot supply any more current (saturation).

The PMT signal is transmitted via a short internal cable to the DOM mainboard. If the voltage level exceeds a certain value, typically set at 0.25 the average SPE amplitude, the DOM records a time-stamped ``hit" and digitizes the signal. This digitized waveform is a sequence of voltage values at successive times, determined by the sampling speed of the particular digitizer. The mainboard features multiple digitizers and gain channels in order to decrease deadtime, increase its dynamic range, and ensure that delayed photons from scattering in the ice are also captured \cite{Abbasi:2008aa}. Recording full waveforms, instead of just time stamps or time-over-threshold, is important for an array in ice because of the complex time structure of photon arrivals.

Most hits detected by a DOM, however, are not from Cherenkov emission in the ice, but rather are ``dark noise" from the PMT and residual radioactivity in the glass pressure sphere. In order to reduce the data volume transmitted up the cable to manageable levels, the DOMs are connected to each other via ``local coincidence" wiring and signal each other when a hit is detected. This local coincidence signal is also forwarded up and down the string to the next-to-nearest neighbor DOMs. Hits with local coincidence indicated from neighbor DOMs are much more likely to be from a particle in the ice, whereas the uncorrelated dark noise hits generally do not have coincidence and can be aggressively compressed before transmission to the surface.

\begin{figure}[ht]
\centering
\includegraphics[width=0.9\linewidth]{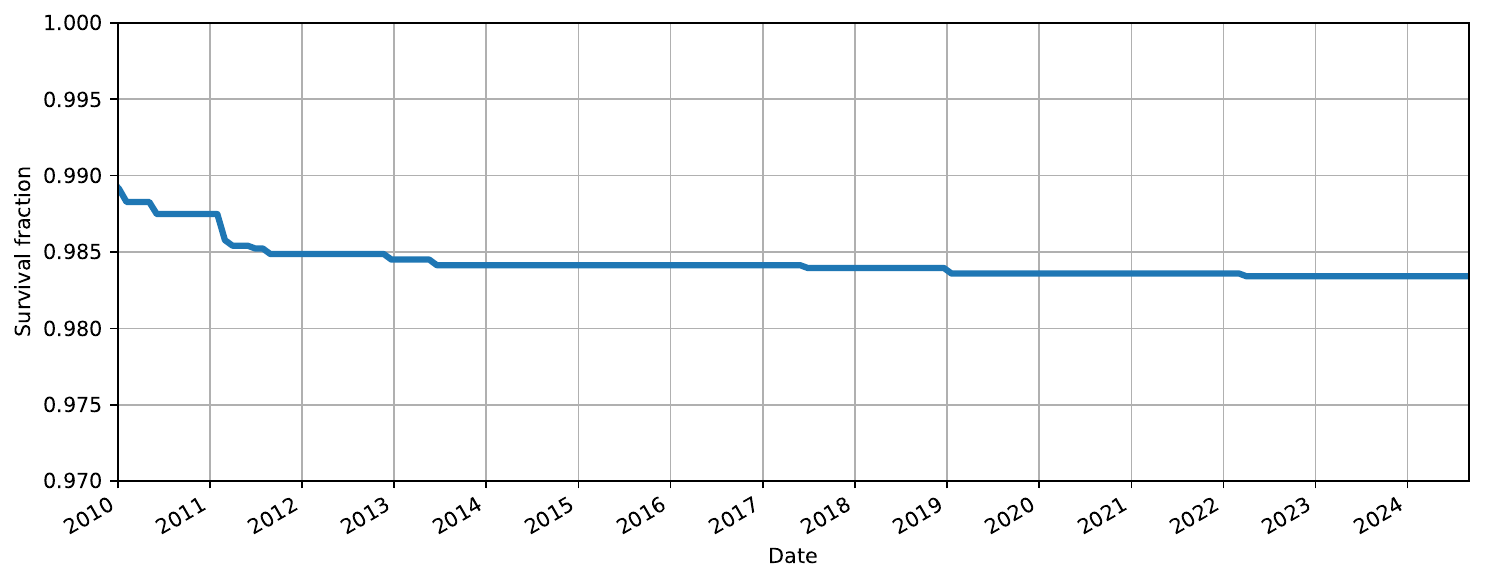}
\caption{DOM survival fraction since 2010. The detector was completed in early 2011.}
\label{fig:dom_survival}
\end{figure}

Because the DOMs cannot be serviced after they are deployed and frozen into the ice sheet, their reliability is key to the long-term operation of the experiment. Each module underwent rigorous testing across a wide temperature range after fabrication and just before deployment. While $\sim1\%$ of the modules failed during deployment or freeze-in, the failure rate after detector completion is less than one DOM per year, and $98.3\%$ of the original 5,484 modules are still operational after 13 years of full-detector operation.  The survival fraction versus time is shown in Fig.~\ref{fig:dom_survival}.

\subsection{Online data systems}\label{daq}

Once the DOM hits are transmitted to the surface, the software data acquisition system (DAQ) forms triggers. Each trigger algorithm is a simple condition based on the pattern of light in the detector in order to collect the data from particle interactions into ``events." The triggers generally operate over time-ordered hits with local coincidence flags and look for $N$ hits within a time window set by the light travel time across the detector (on the order of microseconds). Algorithms can also look for hits within a certain cylindrical area or vertical tracks along a single string. All trigger algorithms are run in parallel, and when a trigger is formed, the triggers are merged. All hits within a readout window before, during, and after the trigger are collected and bundled into an event. The merged trigger rate of IceCube is approximately 2.7 kHz and is dominated by downgoing muons produced in cosmic-ray air showers above the South Pole.

The energy threshold of IceCube using the standard DAQ triggers is too high to detect individual $O(10)$-MeV neutrinos from core-collapse supernova. However, the flux of such neutrinos from a Galactic supernova is large enough to result in a correlated rise in the noise rate of DOMs across the entire detector \cite{IceCube:2011cwc}. To search for such an event, a parallel data acquisition system, the supernova DAQ or SNDAQ, monitors the binned noise rates for all DOMs continuously and searches for significant correlated deviations. In the case of a significant detection, \emph{all} detected photons (pre-trigger hits) from the detector can be saved from a 12.5-day ``hitspool" buffer for subsequent analysis \cite{kopke2018improved}, and an alert is sent to operators and the Supernova Neutrino Early Warning System (SNEWS) \cite{antonioli2004snews}.

Operation of the DAQ and other online systems and monitoring of data quality are facilitated by the IceCube Live software. The experiment control component of IceCube Live normally starts and stops DAQ data-taking runs automatically, but is also used by operators for special operations such as partial-detector runs when maintenance of ICL hardware is required or during calibration activities. IceCube Live monitors the health of the online systems and can page on-site ``winterover" operators in case of a problem. Monitoring data are also collected so that the health of all runs can be assessed by collaboration monitoring shifters. Because of the robustness of the online hardware and software and the automation of detector recovery procedures, IceCube typically is operating more than 99.8\% of the time.

\begin{figure}[ht]
\centering
\includegraphics[width=0.6\linewidth]{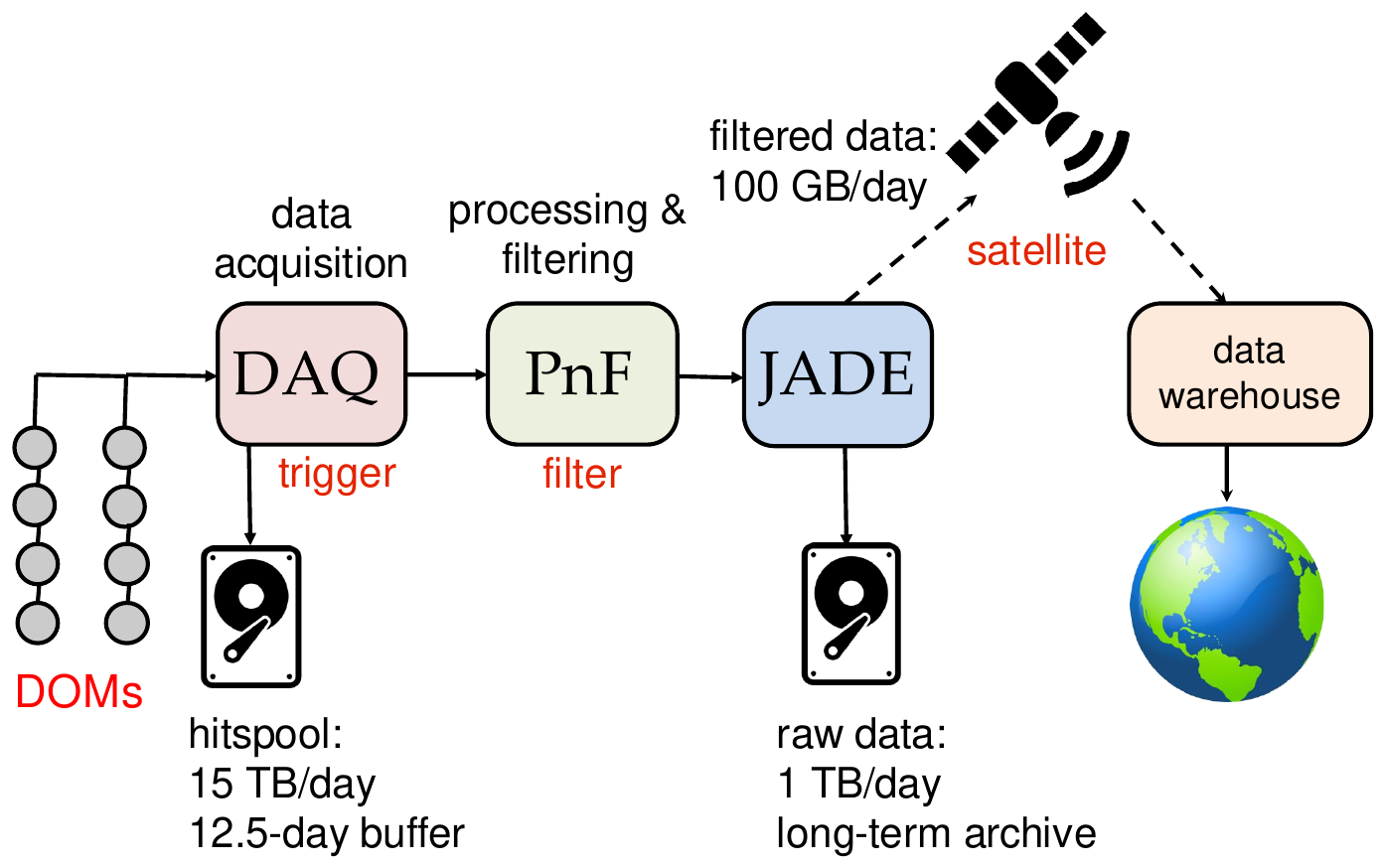}
\caption{\label{fig:data_flow}High-level data flow of the IceCube online systems, from the DOMs to the data warehouse.}
\end{figure}

The triggered events from the DAQ are processed at the South Pole in a small local computing cluster by the processing and filtering (PnF) system. PnF calibrates and performs initial fast directional and energy reconstructions on all events at trigger level. This processing serves two purposes: first, a subset of events can be selected for daily transmission over satellite to the Northern Hemisphere, and second, high-quality astrophysical neutrino candidates can be detected in near real time. Real-time neutrino candidate alerts are then sent to the wider multimessenger astrophysics community for follow-up observation by other telescopes and satellites \cite{Aartsen:2016lmt}. The JADE data handling software manages the storage of both raw data to archival disks, which are flown out of South Pole once per year, as well as the daily satellite data streams that are transferred to the IceCube data warehouse for further processing and analysis. 

The high-level data flow of the online systems from DOMs to the data warehouse is shown in Fig.~\ref{fig:data_flow}. For more detail on the detector instrumentation and online systems, see Ref.~\cite{IceCube:2016zyt}.

\subsection{Calibration}\label{calibration}

In order to measure properties of a neutrino-generated event in the detector, such as direction and deposited energy (via the amount of Cherenkov light detected), the PMT waveforms from the DOMs must first be calibrated from digitizer count and sample number to physical units, i.e., voltage and time. The waveforms then can be integrated (or, in a more sophisticated approach, unfolded using the SPE waveform shape \cite{Aartsen:2013vja}) to obtain charge and the time of each photon arrival at the PMT. The DOMs contain a number of design features to allow them  to calibrate themselves, including a reference charge pulse generation circuit, a precision oscillator for the local time base, and a low-brightness LED on the DOM mainboard that can generate single-photon and larger light pulses. The current waveform of the mainboard LED can also be recorded and analyzed with the detected photons in order to determine the transit time of the PMT and other electronics time delays in the DOM. The dark noise photons are also useful as a calibration source, including in situ measurement of the SPE charge response for each individual DOM \cite{aartsen2020situ}.

\begin{figure}[ht]
\centering
\includegraphics[width=0.4\linewidth]{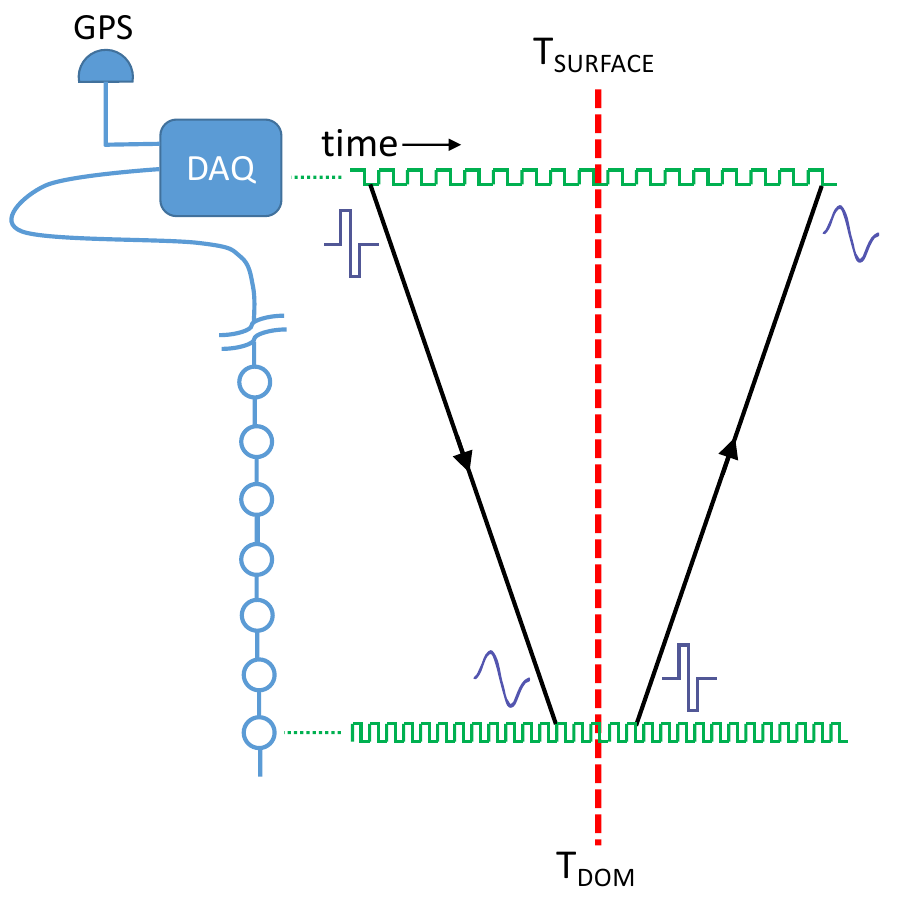}
\caption{\label{fig:rapcal}The RAPCal procedure sends pulses down and up the cable to each DOM every second to translate DOM clock time ($T_{\mathrm{DOM}}$) to surface / UTC time ($T_{\mathrm{surface}}$). } 
\end{figure}

While a precision oscillator in the DOMs provides a stable local time base with which to time stamp waveforms, the DOM clocks are all free-running relative to each other. In order to establish a common time base, the DAQ runs a time calibration procedure from the surface to every DOM in the detector once per second --- RAPCal, or Reciprocal Active Pulsing Calibration. This procedure involves sending a bimodal pulse down the cable to the DOM, digitizing and time-stamping it, and then transmitting a reciprocal pulse back up the cable, where it is again received and time-stamped. Because of the symmetry of the process, the midpoint between the surface time stamps and the DOM time stamps is the same point in time, allowing a translation point from DOM time stamp to surface time (Fig.~\ref{fig:rapcal}).  The surface clocks of the readout hardware in the ICL are all part of a common clock tree sourced from a precision GPS clock. This procedure allows time-stamped DOM hits to be translated to UTC with a relative accuracy of better than 2 ns.

As described previously, the ice sheet forms an integral part of the detector. Understanding the depth- and direction-dependent optical properties of the ice is crucial to mapping the observed timing and charge of detected photons at the DOMs to properties of the neutrino, in particular direction and energy. During dedicated calibration runs, one or more of the 12 LEDs on each DOM's ``flasher board" are flashed, and the emitted light is detected in nearby DOMs. The observed waveforms are compared with GPU-accelerated photon-by-photon simulations of light propagation through the ice sheet, optimizing the scattering and absorption parameters to match observations \cite{abbasi2024situ}. Improvements in the ice modeling have led directly to improvements in angular resolution of the detector and thus sensitivity to neutrino sources.

\subsection{Data processing and analysis}\label{analysis}

Once the online filtered data reach the data warehouse, more computing power is available to run more sophisticated event reconstructions and event selections targeted toward particular physics analyses. IceCube data processing and analysis, including the online filtering, uses the IceTray software. IceTray provides a modular framework within which to process events and subevents, which are stored in ``frames." The trigger level frames from the DAQ (``Q-frames") are generally split into one more physics frames (``P-frames") to separate multiple particle interactions that may have been combined by the DAQ triggers. Event reconstructions run over the extracted PMT pulses (``pulse series"), and reconstructed quantities are saved in processed output files. Physics working groups determine a set of reconstructions and selection criteria to reduce the $O(\mathrm{kHz})$-rate background of downgoing cosmic-ray muons to a more manageable $O(\mathrm{Hz})$-rate subset of events more suitable for defining and tuning further analysis-level processing.

Detailed Monte Carlo simulations are also generated to refine analysis selections that remove background events and to measure the sensitivity of these selections, i.e.,~the fraction of signal events that remain. The simulation software must model all processes leading to the recording of the event, including the neutrino or cosmic-ray interaction, secondary particle production, Cherenkov light emission and propagation in the ice, and relevant aspects of the detector, including the details of the DOM electronics and the DAQ trigger algorithms. The simulation is then processed through the same filtering as the data, allowing measurements of background rejection and signal purity.

Both the data processing and generation and processing of the simulated data require large amounts of computing power. The IceProd software \cite{aartsen2015iceprod} provides a middleware layer that allows processing on a worldwide grid of computing resources, including computing clusters from collaboration institutions and multipurpose scientific computing facilities. A central database manages the metadata needed for each of the simulation or experimental processing data sets.

\section{Physics Topics Addressed by IceCube}\label{physics}

Although the main mission of the AMANDA and IceCube experiments was to reach the scale of detector that makes neutrino astronomy possible, earlier versions of the instrumentation already represented world-leading experiments to look for dark matter. The observation of the annihilation into neutrinos of dark matter particles that have accumulated over time at the center of the Sun would represent a smoking gun for the discovery of dark matter. Establishing upper limits on the cross section for dark matter particles to interact with ordinary matter, the early AMANDA data already challenged the validity of the so-called WIMP miracle. This and other methodologies searching for dark matter remain a priority to date \cite{aartsen2017first, Aartsen:2017ulx, abbasi2022search}.

Taking data continuously with essentially no deadtime, IceCube also monitors our Galaxy for the explosion of a supernova. The observation of roughly a million electron antineutrinos from an event at the most probable distance of $10\,\rm kpc$ will provide a detailed movie of the time evolution of the explosion. In the case of such an event, the data collected in these seconds may very well provide more information for astrophysics and particle physics than produced by the instrument before or after.

With the addition of the DeepCore strings at the center of the detector, the increased photocathode coverage of two megatons of clear deep ice resulted in an instrument with a 5 GeV neutrino threshold. The high-statistics atmospheric neutrino data collected with DeepCore allowed us to produce 3-neutrino flavor oscillation analyses that are competitive with those performed with accelerator-based experiments \cite{icecube2024measurement}. In such analyses, the zenith angle of the arriving atmospheric neutrino serves as a proxy for the oscillation baseline, up to the diameter of the Earth. These analyses are unique because they are directly sensitive to both the disappearance of muon neutrinos and their reappearance as tau neutrinos in the atmospheric neutrino beam.

With some neutrinos exceeding the energy of those produced at Fermilab by a factor of a million, the extraordinary opportunity to search for physics beyond the Standard Model is evident. At the same time, we can perform tests of its fundamental symmetries, such as Lorenz invariance and the equivalence principle, that are often world leading~\cite{Aartsen:2017ibm}.

In addition to this broad range of physics topics, IceCube has accumulated by now an extensive record of studying the glaciology of the Antarctic ice sheet \cite{abbasi2024situ} and has an active program performing tomography of the structure of the geology of the Earth using neutrinos.

We may anticipate using IceCube as a tool for science in ways that have not been imagined yet.

\section{Complementarity with Other Large-scale Neutrino Detectors}\label{compl}

The AMANDA and IceCube experiments built on the pioneering research of the DUMAND~\cite{Babson:1989yy} collaboration, which initiated the construction of a neutrino telescope in the deep ocean water off the coast of Hawaii. After that effort was discontinued, several neutrino telescope projects were initiated in the Mediterranean Sea and in Lake Baikal in the 1990s~\cite{BAIKAL:1997iok,Aggouras:2005bg,Aguilar:2006rm,Migneco:2008zz}. In 2008, the construction of the ANTARES detector off the coast of France was completed. It demonstrated the feasibility of neutrino detection in the deep sea and has provided a wealth of technical experience and design solutions for deep-sea components. 

An international collaboration has started construction of a cubic kilometer neutrino telescope in the Mediterranean Sea, KM3NeT~\cite{Adrian-Martinez:2016fdl}, with a detector volume anticipated to be sufficient for initial observations of the neutrino sky; see e.g.,~\cite{Roberts:1992,Gaisser1995}. They developed a digital optical module that incorporates 31 3-inch photomultipliers instead of one large photomultiplier tube. The advantages are a tripling of the photocathode area per optical module, a segmentation of the photocathode allowing for a clean identification of coincident Cherenkov photons, some directional sensitivity, and a reduction of the overall number of penetrators and connectors, which are expensive and prone to failure. KM3NeT in its second phase~\cite{Adrian-Martinez:2016fdl} will consist of two units for astrophysical neutrino observations, each consisting of 115 strings carrying more than 2,000 optical modules. 

A parallel effort is underway in Lake Baikal with the construction of the deep underwater neutrino telescope Baikal-GVD (Gigaton Volume Detector)~\cite{Avrorin:2015wba}. The first GVD cluster was upgraded in the spring of 2016 to its final size: 288 optical modules, a geometry of 120 meters in diameter and 525 meters high, and an instrumented volume of 6 Mton. Each of the eight strings consists of three sections with 12 optical modules. At this time, 7 of the 14 clusters have been deployed, reaching a sensitivity close to the diffuse cosmic neutrino flux observed by IceCube. 

Projects in China aim for larger detectors, two planning for a $10~\rm km^2$-scale design ~\cite{Ye:2022vbk,Zhang:2024slv}, similar to the next-generation IceCube detector, and one envisioning a larger volume of $\SI{30}\km^3$~\cite{Huang:2023mzt}.

\section{Main Results, Open Questions, and Future Development}\label{future}

Imagined more than half a century ago~\cite{Spiering:2012xe}, high-energy neutrino astronomy became a reality when the IceCube project transformed a cubic kilometer of transparent natural Antarctic ice one mile below the geographic South Pole into the largest particle detector ever built~\cite{IceCube:2016zyt}.
It discovered neutrinos in the TeV to PeV energy range originating beyond our Galaxy, opening a new window for astronomy~\cite{Aartsen:2013jdh,IceCube:2014stg}. It turned out that the observed energy density of neutrinos in the extreme universe exceeds the energy in gamma rays observed by Fermi~\cite{Fang:2022trf}; it even outshines the neutrino flux from the nearby sources in our own Galaxy~\cite{IceCube:2023ame}.
The Galactic plane only appears as a faint glow at the ten percent level of the extragalactic flux, in sharp contrast with all other wavelengths of light where the Milky Way is the dominant feature in the sky.
This observation implies the existence of sources of high-energy neutrinos in other galaxies that are not present in our own~\cite{Fang:2023azx}. A reasonable explanation may be the absence of activity of the central black hole in our own Galaxy in the last few million years.

After accumulating a decade of data with a detector with gradually enhanced sensitivity, the first high-energy neutrino sources emerged in the neutrino sky: the active galaxies NGC~1068, NGC~4151, PKS~1424+240~\cite{IceCube:2022der,IceCube:2023jds}, and TXS~0506+056~\cite{IceCube:2018dnn,IceCube:2018cha}. IceCube even found evidence for neutrino emission from the Circinus active galaxy despite the detector's reduced sensitivity in the Southern Hemisphere~\cite{Huang:2023mzt}.
The observations point to the acceleration of protons and the production of neutrinos in the obscured dense cores surrounding the supermassive black holes at their center, typically within a distance of only $\sim10$ Schwarzschild radii~\cite{IceCube:2022der}.
TXS~0506+056 had previously been identified as a neutrino source in a multimessenger campaign triggered by a neutrino of $\SI{290}\TeV$ energy, IC170922, and by the independent observation of a neutrino burst from the same source in archival IceCube data in 2014~\cite{IceCube:2018dnn,IceCube:2018cha}.

With indications that neutrinos originate in a vicinity of the central black hole of active galaxies, which can only be reached by radio telescopes, neutrino astronomy represents an extraordinary opportunity for discovery.
This includes resolving the century-old problem of where cosmic rays originate and how they reach their phenomenal energies. Because neutrinos originate from the decay of charged pions, only sources accelerating cosmic rays that interact with radiation fields or gas surrounding the accelerator to produce pions can be neutrino sources.
With energies exceeding those produced with earthbound accelerator beams by a factor of one million, cosmic neutrinos also present a remarkable opportunity for studying the neutrinos themselves.

Despite its exceptional discovery potential, neutrino astronomy eventually faces a lack of statistics for future progress, not only in neutrinos but also in identified sources, a handful in the last decade.
A weakness in the current status represents a future challenge and opportunity.
As previously discussed, the sensitivity of IceCube has recently been significantly enhanced by progress in the characterization of the optics of ice, and by exploiting the rapid progress in machine learning to collect larger data samples that are reconstructed in neutrinos with improved energy and angular resolution.
The observations of NGC~1068 and the Galactic plane directly resulted from these improvements.

Efforts to improve the performance of IceCube continue, with an additional seven strings of newly designed optical modules and calibration instruments planned for deployment in 2025--26. The IceCube Upgrade will lower the energy threshold of the detector via a dense infill of new instrumentation and provide a high-statistics sample of atmospheric neutrinos for precision neutrino oscillation measurements \cite{ishihara2019icecube}. New calibration light sources and cameras will allow continued improvements in the determination of the optical properties of the ice, including  characterization of the refrozen ``bubble column" formed when the drill holes refreeze. These improvements are expected to increase the sensitivity of the entire historical IceCube data set.

However, progress in the science will inevitably require higher statistics of neutrinos with superior angular resolution. As a next step, IceCube has produced a technical design for a next-generation detector, IceCube-Gen2, instrumenting more than $8\rm\,km^3$ of glacial ice at the South Pole, capitalizing on the large absorption length of Cherenkov light revealed by the construction of the first-generation detectors~\cite{IceCube-Gen2:2020qha}. Exploiting the extremely long photon absorption lengths in the deep Antarctic ice, the spacing between strings of light sensors will be increased from 125 to close to 250 meters without significant loss of performance of the instrument at TeV energies and above. This has been verified by analyzing IceCube data with half of the strings removed. The instrumented volume can thus grow by close to one order of magnitude while keeping the instrumentation and its cost at the level of IceCube. The new facility will increase the rates of cosmic events from hundreds to thousands over several years. The superior angular resolution of the longer muon tracks will allow for the identification of the many cosmic neutrino sources currently seen close to the $3\sigma$ level in the present data. The design for such an instrument exists~\cite{IceCube-Gen2:2020qha}, and its construction has been strongly endorsed by both the astronomy~\cite{NAP26141} and particle physics communities~\cite{P5:2023wyd}.


\section{Conclusions}
\label{sec:conclusions}

In the last decade, IceCube has established the viability of neutrino astronomy. Its observation of a diffuse extragalactic neutrino flux that outshines the flux observed of the highest energy gamma rays also produced the field's first surprise: it dominates the flux from nearby sources in our own Galaxy by about an order of magnitude. With the identification of the first neutrino sources, IceCube has demonstrated that, with multimessenger astronomy, we have created the tools to identify the origin of the highest energy particles reaching us from the universe.

\begin{ack}[Acknowledgments]%
 We thank our IceCube colleagues for the many years of dedicated collaborations that make this article possible. We thank Chris Wendt for the waveform samples shown in Fig.~\ref{fig:pmt_waveforms}. We thank the National Science Foundation for ongoing support for IceCube construction, management and operations, and data analysis. This material is based upon work supported by the National Science Foundation under Award Nos. 1719277, 2042807, and 2209445.
\end{ack}

\bibliographystyle{elsarticle-num} 
\bibliography{reference}

\end{document}